\documentclass[runningheads]{llncs}

\usepackage{amsmath}
\usepackage{amsfonts}
\usepackage{amssymb}
\usepackage{graphicx}
\usepackage[colorlinks]{hyperref}
\usepackage[switch]{lineno}

\usepackage[disable]{todonotes}

\title{Assertion-Based Approaches to Auditing Complex Elections, with
Application to Party-List Proportional Elections\thanks{Published in:
Krimmer R. et al.\ (eds) Electronic Voting. E-Vote-ID 2021. LNCS 12900.
Springer, Cham. \url{https://doi.org/10.1007/978-3-030-86942-7_4}}}

\titlerunning{Assertion-Based Approaches to Auditing Complex Elections}

\author{
Michelle Blom     \inst{1}   \orcidID{0000-0002-0459-9917}  \and
Jurlind Budurushi \inst{2}   \orcidID{0000-0002-6732-4400}  \and
Ronald L. Rivest  \inst{3}   \orcidID{0000-0002-7105-3690}  \and
Philip B. Stark   \inst{4}   \orcidID{0000-0002-3771-9604}  \and
Peter J. Stuckey  \inst{5}   \orcidID{0000-0003-2186-0459}  \and
Vanessa Teague    \inst{6}   \orcidID{0000-0003-2648-2565}  \and
Damjan Vukcevic   \inst{7,8} \orcidID{0000-0001-7780-9586}}

\authorrunning{Blom, Budurushi, Rivest, Stark, Stuckey, Teague, Vukcevic}

\institute{
School of Computing and Information Systems, University of Melbourne,
Parkville, Australia \\
\email{michelle.blom@unimelb.edu.au}
\and
Cloudical Deutschland GmbH, Berlin, Germany \\
\email{jurlind.budurushi@cloudical.io}
\and
Computer Science and Artificial Intelligence Laboratory, Massachusetts
Institute of Technology, Cambridge, MA, USA
\and
Department of Statistics, University of California, Berkeley, CA, USA
\and
Department of Data Science and AI, Monash University, Clayton, Australia
\and
Thinking Cybersecurity Pty. Ltd., Melbourne, Australia
\and
School of Mathematics and Statistics, University of Melbourne, Parkville,
Australia
\and
Melbourne Integrative Genomics, University of Melbourne, Parkville,
Australia \\
\email{damjan.vukcevic@unimelb.edu.au}}

\begin{document}

\maketitle

\begin{abstract}
Risk-limiting audits (RLAs), an ingredient in evidence-based elections, are
increasingly common. They are a rigorous statistical means of ensuring that
electoral results are correct, usually without having to perform an expensive
full recount---at the cost of some controlled probability of error. A recently
developed approach for conducting RLAs, SHANGRLA, provides a flexible framework
that can encompass a wide variety of social choice functions and audit
strategies. Its flexibility comes from reducing sufficient conditions for
outcomes to be correct to canonical `assertions' that have a simple
mathematical form.

Assertions have been developed for auditing various social choice functions
including plurality, multi-winner plurality, super-majority, Hamiltonian
methods, and instant runoff voting. However, there is no systematic approach
to building assertions. Here, we show that assertions with \emph{linear}
dependence on transformations of the votes can easily be transformed to
canonical form for SHANGRLA. We illustrate the approach by constructing
assertions for party-list elections such as Hamiltonian free list elections and
elections using the D'Hondt method, expanding the set of social choice
functions to which SHANGRLA applies directly.

\end{abstract}


\section{Introduction}
\label{sec:intro}

\emph{Risk-limiting audits} (RLAs) test reported election outcomes
statistically by manually inspecting random samples of paper ballots. An RLA
terminates either by endorsing the reported outcome or by proceeding to a full
manual count if the evidence is inconclusive. The outcome according to the
full count corrects the reported outcome if they differ. The \emph{risk limit}
is an upper bound on the probability that a wrong election outcome will not be
corrected---this is set in advance, typically between 1\% and 10\%.

SHANGRLA~\cite{shangrla} is a general framework for conducting RLAs of a wide
variety of social choice functions.\footnote{%
Any social choice function that is a \emph{scoring rule}---that assigns
`points' to candidates on each ballot, sums the points across ballots, and
declares the winner(s) to be the candidate(s) with the most `points'---can be
audited using SHANGRLA, as can some social choice functions that are not
scoring rules, such as super-majority and IRV.}
SHANGRLA involves reducing the correctness of a reported outcome to the truth
of a set $\mathcal{A}$ of quantitative \emph{assertions} about the set of
validly cast ballots, which can then be tested using statistical methods. The
assertions are either true or false depending on the votes on the ballots. If
all the assertions are true, the reported outcome is correct.

This paper shows how to use the SHANGRLA RLA method to audit some complex
social choice functions not addressed in the SHANGRLA paper. We give a recipe
for translating sufficient conditions for a reported outcome to be correct into
canonical form for SHANGRLA, when those conditions are the intersection of a
set of linear inequalities involving transformations of the votes on each
ballot. We focus on European-style party-list proportional representation
elections, with the German state of Hesse as a case study.

\subsection{Assertion-based auditing: Properties and challenges}

For some social choice functions, the reduction to assertions is obvious. For
instance, in plurality (first-past-the-post) elections, common in the United
States, Alice won the election if and only if Alice's tally was higher than
that of each of the other $n-1$ candidates (where $n$ is the total number of
candidates). That set of $n-1$ assertions is clearly a set of linear
inequalities among the vote totals for the $n$ candidates.

In general, assertions involve not only the votes but also the reported
results---the reported outcome and possibly the voting system's interpretation
of individual ballots (CVRs) or tallies of groups of ballots.

SHANGRLA~\cite[Sec 2.5]{shangrla} shows how to make assorters for any `scoring
rule' (e.g.\ Borda, STAR-voting, and any weighted scheme). For more complex
social choice functions, constructing sufficient sets of assertions may be much
less obvious. Blom {\it et al.}~\cite{blom2019raire} use a heuristic method,
RAIRE, to derive assertions for Instant Runoff Voting (IRV) from the CVRs.
RAIRE allows the RLA to test an IRV outcome---the claim that Alice
won---without checking the entire IRV elimination. RAIRE's assertions are
\emph{sufficient}: if all of the assertions in $\mathcal{A}$ are true, then the
announced election outcome is correct. However, the set of assertions might not
be necessary---even if one of the assertions in $\mathcal{A}$ is false, Alice
may still have won, but for reasons not checked by the audit.

A social choice function might be expensive to audit for two different reasons:
it might require a very large sample for reasonable confidence, even when there
are no errors (for instance, if it tends to produce small margins in practice);
alternatively, it might be so complex that it is difficult to generate
assertions that are sufficient to prove the reported election outcome is
correct. Pilots and simulations suggest that IRV elections do not have small
margins any more often than first-past-the-post elections. Hence IRV is
feasible to audit in both senses.

Below, the sets of assertions we consider are \emph{conjunctive}: the election
outcome is correct if all the assertions in $\mathcal{A}$ are true. Although it
is possible to imagine an audit method that tests more complex logical
structures (for example, the announced outcome is correct if either all the
assertions in $\mathcal{A}_1$ or all the assertions in $\mathcal{A}_2$ are
true), this is not currently part of the SHANGRLA framework.

\paragraph{Summary:} An audit designer must devise a set $\mathcal{A}$ of
assertions.
\begin{itemize}
\item $\mathcal{A}$ generally depends on the social choice function and the
    reported electoral outcome, and may also depend on the CVRs, vote
    subtotals, or other data generated by the voting system.

\item If every assertion in $\mathcal{A}$ is true, then the announced electoral
    result is correct.

\item The announced electoral result may be correct even if not every assertion
    in $\mathcal{A}$ is true.
\end{itemize}
SHANGRLA relies on expressing assertions in terms of \emph{assorters}.

\subsection{Assorters}

The statistical part of SHANGRLA is agnostic about the social choice function.
It simply takes a collection of sets of numbers that are zero or greater (with
a known upper bound), and decides whether to reject the hypothesis that the
mean of each set is less than or equal to $1/2$---this is the \emph{assorter
null hypothesis}.

An \emph{assorter} for some assertion $A \in \mathcal{A}$ assigns a nonnegative
value to each ballot, depending on the selections the voter made on the ballot
and possibly other information (e.g.\ reported vote totals or CVRs). The
assertion is true iff the mean of the assorter (over all ballots) is greater
than 1/2. Generally, ballots that support the assertion score higher than
$1/2$, ballots that cast doubt on it score less than $1/2$, and neutral ballots
score exactly $1/2$. For example, in a simple first-past-the-post contest, $A$
might assert that Alice's tally is higher than Bob's. The corresponding
assorter would assign 1 to a ballot if it has a vote for Alice, 0 if it has a
vote for Bob, and $1/2$ if it has a no valid vote or a vote for some other
candidate.

The audit designer's first job is to generate a set $\mathcal{A}$ of assertions
which, if all true, imply that the announced electoral outcome (the winner or
winners) is correct. Then they need to express each $A \in \mathcal{A}$ using
an assorter. Finally, they need to test the hypothesis that any assorter mean
is less than or equal to 1/2. If all those hypotheses are rejected, the audit
concludes that the reported outcome is correct. The chance this conclusion is
erroneous is at most the risk limit.

\autoref{sec:assertionsAndAssorters} gives a more precise definition of an
assorter and a general technique for transforming linear assertions into
assorters.

\subsection{Risk-limiting audits using SHANGRLA: Pulling it all together}

An overview of the workflow for a sequential SHANGRLA RLA is:
\begin{enumerate}
\item Generate a set of assertions.
\item Express the assertions using assorters.
\item Test every assertion in $\mathcal{A}$, in parallel:
    \begin{enumerate}
        \item Retrieve a ballot or set of ballots selected at random.
        \item Apply each assorter to every retrieved ballot.
        \item For each assertion in $\mathcal{A}$, test its corresponding
            assorter null hypothesis (i.e.\ that the assorter mean is
            $\leqslant 1/2$) using a sequentially valid test.\footnote{%
            It can be more efficient to sample ballots in `rounds' rather than
            singly; SHANGRLA can accommodate any valid test of the assorter
            nulls.}
        \item If the assorter null is rejected for $A \in \mathcal{A}$, remove
            $A$ from $\mathcal{A}$.
        \item If $\mathcal{A}$ is empty (i.e.\ all of the null hypotheses have
            been rejected), stop the audit and certify the electoral outcome.
        \item Otherwise, continue to sample more ballots.
        \item At any time, the auditor can decide to `cut to the chase' and
            conduct a full hand count: anything that increases the chance of
            conducting a full hand count cannot increase the risk.
    \end{enumerate}
\end{enumerate}
As with any RLA, the audit may not confirm the reported result (for example,
that Alice's tally is the highest) even if all assertions are true (Alice's
tally may actually be higher than Bob's, but the audit may not gather enough
evidence to conclude so). This may happen because there are many tabulation
errors or because one or more margins are small. When the audit proceeds to a
full hand count, its result replaces the reported outcome if the two differ.

Conversely, the audit may mistakenly confirm the result even if the announced
result is wrong. The probability of this kind of failure is not more than the
\emph{risk limit}. This is a parameter to SHANGRLA; setting it to a smaller
value generally entails examining more ballots.

\subsection{Party-list proportional representation contests}

Party-list proportional representation contests allocate seats in a parliament
(or delegates to an assembly) in proportion to the entities' popularity within
the electorate. The first step is (usually) rounding the party's fraction down
to the nearest integer number of seats. Complexity arises from rounding, when
the fractions determined by voters do not exactly match integer numbers of
seats. \emph{Largest Remainder Methods}, also called Hamiltonian methods,
successively allocate leftover seats to the entities with the largest
fractional parts until all seats are allocated. \emph{Highest Averages
Methods}, such as the D'Hondt method (also called Jefferson's method), weight
this extra allocation by divisors involving a fraction of the seats already
allocated to that party---they are hence more likely to allocate the leftover
seats to small parties. The Sainte-Lagu\"{e} method (also called Webster's
method) is mathematically similar but its divisors penalise large parties even
more.\footnote{%
Another source of complexity is the opportunity for voters to select, exclude,
or prioritise individual candidates within the party.}

\subsection{Related work and our contribution}

Blom {\it et al.}~\cite{hamiltonian} showed how to construct a SHANGRLA RLA for
preferential Hamiltonian elections with a viability threshold, applicable to
many US primaries. Stark and Teague~\cite{stark2014verifiable} showed how to
construct an RLA for highest averages party-list proportional representation
elections. Their method was not directly based on assertions and assorters,
but it reduces the correctness of the reported seat allocation to a collection
of two-entity plurality contests, for which it is straightforward to construct
assorters, as we show below.

This paper shows how to extend SHANGRLA to additional social choice functions.
We use party-list proportional representation elections as an example, showing
how the assorter from~\cite{hamiltonian} can be derived as a special case of
the solution for more general Hamiltonian elections. We have simulated the
audit on election data from the German state of Hesse; results are shown in
\autoref{sec:HesseExample}. Auditing the allocation of integer portions of
seats involves inspecting a reasonable number of ballots, but the correctness
of the allocations based on the fractional remainders and the correctness of
the particular candidates who receive seats within each party generally involve
very small margins, which in turn require large audit sample sizes.  We also
show how to apply the construction to highest averages methods such as D'Hondt
and Sainte-Lagu\"{e}. Our contributions are:
\begin{itemize}
\item A guide to developing assertions and their corresponding SHANGRLA
    assorters, so that audits for contest types that are not already supplied
    can be derived, when correctness can be expressed as the intersection of a
    set of linear inequalities (\autoref{sec:GeneratingAssorters}).
\item New SHANGRLA-based methods for auditing largest remainder methods that
    allow individual candidate selection (no audit method was previously known
    for this variant of largest remainder method) (\autoref{sec:pairdiff}).
\item Simulations to estimate the average sample sizes of these new methods in
    the German state of Hesse (\autoref{sec:HesseExample}).
\item SHANGRLA assorters for highest averages methods (RLAs for these methods
    were already known, but had not been expressed as assorters).
    (\autoref{sec:dHondtExample}).
\end{itemize}


\section{Preliminaries}
\label{sec:assertionsAndAssorters}

\newcommand{\election}{\mathcal{L}}
\newcommand{\prop}[1]{p_{#1}}
\newcommand{\tally}[1]{T_{#1}}
\newcommand{\ballot}[1]{b_{#1}}
\newcommand{\ballottotal}{b_T}
\newcommand{\maxvote}{m_\election}
\newcommand{\totalvote}{T_\election}
\newcommand{\maxindvote}{m}
\newcommand{\seats}{S}
\newcommand{\entities}{E}

\subsection{Nomenclature and notation for assertion-based election audits}

An election contest is decided by a set of `ground truth' ballots $\election$
(of cardinality $|\election|$). Many social choice functions are used in
political elections. Some yield a single winner; others multiple winners. Some
only allow voters to express a single preference; others allow voters to select
or rank multiple candidates or parties.

Here, we focus on elections that allow voters to select (but not rank) one or
more `entities,' which could be candidates or parties.\footnote{%
Below, in discussing assorters, we use the term `entity' more abstractly. For
instance, when voters may rank a subset of entities, the assorters may
translate ranks into scoring functions in a nonlinear manner, as
in~\cite{blom2019raire}---we do not detail that case here.}

Let $\seats$ be the number of `seats' (positions) to be filled in the contest,
of which $a_e$ were awarded to entity $e$. Each ballot might represent a
single vote for an entity, or multiple votes for multiple entities. Important
quantities for individual ballots $b \in \election$ include:
\begin{itemize}
\item $\maxindvote$, the maximum permitted number of votes for any entity.
\item $\maxvote$, the maximum permitted number of votes in total (across all
    entities).
\item $\ballot{e}$, the total number of (valid) votes for entity $e$ on the the
    ballot.
\item $\ballottotal := \sum_{e \in \entities} \ballot{e}$, the total number of
    (valid) votes on the ballot.
\end{itemize}
Any of these may be greater than one, depending on the social choice function.
Validity requires $\ballot{e} \leqslant \maxindvote$ and $\ballottotal
\leqslant \maxvote$. If ballot $b$ does not contain the contest in question or
is deemed invalid, $\ballot{e} := 0$ for all entities $\entities$, and
$\ballottotal := 0$.

Important quantities for the set $\election$ of ballots include:
\begin{itemize}
\item $\tally{e} = \sum_{b \in \election} \ballot{e}$, the \emph{tally} of
    votes for entity $e$.
\item $\totalvote = \sum_{e \in \entities} \tally{e}$, the total number of
    valid votes in the contest.
\item $\prop{e} = \tally{e} / \totalvote$, the \emph{proportion} of votes for
    entity $e$.
\end{itemize}

\subsection{Assertion-based auditing: Definitions}

Here we formalize assertion-based auditing sketched in \autoref{sec:intro} and
introduce the relevant mathematical notation. An \emph{assorter} $h$ is a
function that assigns a non-negative number to each ballot depending on the
votes reflected on the ballot and other election data (e.g.\ the reported
outcome, the set of CVRs, or the CVR for that ballot). Each assertion in the
audit is equivalent to `the average value of the assorter for all the cast
ballots is greater than 1/2.' In turn, each assertion is checked by testing the
complementary null hypothesis that the average is less than or equal to 1/2.
If all the complementary null hypotheses are false, the reported outcome of
every contest under audit is correct.
\begin{definition}
An \emph{assertion} is a statement $A$ about the set of paper ballots
$\election$ of the contest. An \emph{assorter} for assertion $A$ is a function
$h_A$ that maps selections on a ballot $b$ to $[0, M]$ for some known constant
$M>0$, such that assertion $A$ holds for $\election$ iff $\bar{h}_A > 1/2$
where $\bar{h}_A$ is the average value of $h_A$ over all $b \in \election$.
\end{definition}
A set $\mathcal{A}$ of assertions is \emph{sufficient} if their conjunction
implies that the reported electoral outcome is correct.

\subsection{Example assertions and assorters}

\begin{example}
\label{ex:fptp}
\emph{First-past-the-post voting.} Consider a simple first-past-the-post
contest, where the winner $w$ is the candidate with the most votes and each
valid ballot records a vote for a single candidate. The result is correct if
the assertions $\prop{w} > \prop{\ell}$ for each losing candidate $\ell$ all
hold.

We can build an assorter $h$ for the assertion $\prop{w} > \prop{\ell}$ as
follows \cite{shangrla}:
\[
h(b) :=
    \begin{cases}
             1     & \ballot{w} = 1 \text{ and } \ballot{\ell} = 0, \\
             0     & \ballot{w} = 0 \text{ and } \ballot{\ell} = 1, \\
       \frac{1}{2} & \text{otherwise.}
    \end{cases}
\]
\end{example}

\begin{example}
\emph{Majority contests.}
Consider a simple majority contest, where the winner is the candidate $w$
achieving over 50\% of the votes, assuming again each valid ballot holds a
single vote (if there is no winner, a runoff election is held). The result can
be verified by the assertion $\prop{w} > 1/2$.

We can build an assorter $h$ for the more general assertion $\prop{w} > t$ as
follows \cite{shangrla}:
\[
h(b) :=
\begin{cases}
  \frac{1}{2t} & \ballot{w} = 1 \text{ and } \ballot{\ell} = 0,
                                                        \forall \ell \ne w, \\
        0      & \ballot{w} = 0 \text{ and } \ballot{\ell} = 1
                                       \text{ for exactly one } \ell \ne w, \\
  \frac{1}{2}  & \text{invalid ballot.}
\end{cases}
\]
\end{example}

%


\section{Creating assorters from assertions}
\label{sec:GeneratingAssorters}

In this section we show how to transform generic linear assertions, i.e.\
inequalities of the form $\sum_{b \in \election} \sum_{e \in E} a_e \ballot{e}
> c$, into canonical assertions using assorters as required by SHANGRLA. There
are three steps:
\begin{enumerate}
\item Construct a set of linear assertions that imply the correctness of the
    outcome.\footnote{%
    Constructing such a set is outside the scope of this paper; we suspect there
    is no general method. Moreover, there may be social choice functions for
    which there is no such set.}
\item Determine a `proto-assorter' based on this assertion.
\item Construct an assorter from the proto-assorter via an affine
    transformation.
\end{enumerate}
We work with social choice functions where each valid ballot can contribute a
non-negative (zero or more) number of `votes' or `points' to various tallies
(we refer to these as \emph{votes} henceforth). For example, in plurality
voting we have a tally for each candidate and each ballot contributes a vote of
1 to the tally of a single candidate and a vote of 0 to all other candidates'
tallies. The tallies can represent candidates, groups of candidates, political
parties, or possibly some more abstract groupings of candidates as might be
necessary to describe an assertion (see below); we refer to them generically as
\emph{entities}.

Let the various tallies of interest be $\tally{1}, \tally{2}, \dots, \tally{m}$
for $m$ different entities. These represent the total count of the votes
across all valid ballots.

A \emph{linear assertion} is a statement of the form
\[ a_1 \tally{1} + a_2 \tally{2} + \dots + a_m \tally{m} > 0 \]
for some constants $a_1, \dots, a_m$.

Each assertion makes a claim about the ballots, to be tested by the audit. For
most social choice functions, the assertions are about proportions rather than
tallies. Typically these proportions are of the total number of valid votes,
$\totalvote$, in which case we can restate the assertion in terms of tallies by
multiplying through by $\totalvote$.

For example, a pairwise majority assertion is usually written as $\prop{A} >
\prop{B}$, stating that candidate $A$ got a larger proportion of the valid
votes than candidate $B$. We can write this in linear form as follows. Let
$\tally{A}$ and $\tally{B}$ be the tallies of votes in favour of candidates $A$
and $B$ respectively. Then:
\begin{align*}
         \prop{A}              &>        \prop{B}               \\
  \frac{\tally{A}}{\totalvote} &> \frac{\tally{B}}{\totalvote}  \\
        \tally{A}              &>       \tally{B}               \\
        \tally{A} - \tally{B}  &>       0.
\end{align*}

Another example is a super/sub-majority assertion, $\prop{A} > t$, for some
threshold $t$. We can write this in linear form similar to above, as follows:
\begin{align*}
         \prop{A}                   &> t                \\
  \frac{\tally{A}}{\totalvote}      &> t                \\
        \tally{A}                   &> t \, \totalvote  \\
        \tally{A} - t \, \totalvote &> 0.
\end{align*}

For a given linear assertion, we define the following function on ballots,
which we call a \emph{proto-assorter}:
\[ g(b) = a_1 b_1 + a_2 b_2 + \dots + a_m b_m, \]
where $b$ is a given ballot, and $b_1, b_2, \dots, b_m$ are the votes
contributed by that ballot to the tallies $T_1, T_2, \dots, T_m$
respectively.\footnote{%
Note that $g(b) = 0$ for any invalid ballot $b$, based on previous
definitions.}

Summing this function across all ballots, $\sum_b g(b)$, gives the left-hand
side of the linear assertion. Thus, the linear assertion is true iff $\sum
g(b) > 0$. The same property holds for the average across ballots, $\bar{g} =
|\election|^{-1} \sum g(b)$; the linear assertion is true iff $\bar{g} > 0$.

To obtain an assorter in canonical form, we apply an affine transformation to
$g$ such that it never takes negative values and also so that comparing its
average value to $1/2$ determines the truth of the assertion. One such
transformation is
\begin{equation}
\label{eq:generalAssorter}
  h(b) = c \cdot g(b) + 1/2
\end{equation}
for some constant $c$.\footnote{Note that $h(b) = 1/2$ if ballot $b$ has no
valid vote in the contest.} There are many ways to choose $c$. We present two
here.
First, we determine a lower bound for the proto-assorter, a value $a$ such that
$g(b) \geqslant a$ for all $b$.\footnote{%
If the votes $b_j$ are bounded above by $s$ and below by zero, then a bound
(not necessarily the sharpest) on $g$ is given by taking just the votes that
contribute negative values to $g$, setting all of those votes to $s$, and
setting the other votes to 0:
\[ a = \sum_{j : a_j < 0} a_j s. \]}
Note that $a < 0$ in all interesting cases: if not, the assertion would be
trivially true ($\bar{g} > 0$) or trivially false ($\bar{g} \equiv 0$, with
$a_j = 0$ for all $j$).
If $a \geqslant -1/2$, simply setting $c = 1$ produces an assorter:
we have $h \geqslant 0$, and $\bar{h} > 1/2$ iff $\bar{g} > 0$.
Otherwise, we can choose $c = -1/(2a)$, giving
\begin{equation}
\label{eq:assorter}
  h(b) = \frac{g(b) - a}{-2a}.
\end{equation}
(See \cite[Sec. 2.5]{shangrla}.)
To see that $h(b)$ is an assorter, first note that $h(b) \geqslant 0$ since the
numerator is always non-negative and the denominator is positive.
Also, the sum and mean across all ballots are, respectively:
\begin{align*}
    \sum_b h(b) &= -\frac{1}{2a} \sum_b g(b) + \frac{|\election|}{2}  \\
      \bar{h}   &= -\frac{1}{2a}   \bar{g}   + \frac{1}{2}.
\end{align*}
Therefore, $\bar{h} > 1/2$ iff $\bar{g} > 0$.

\subsection{Example: Pairwise difference assorter}
\label{sec:pairdiff}

To illustrate the approach, we will now create an assorter for a fairly complex
assertion for quite complicated ballots. We consider a contest where each
ballot can have multiple votes for multiple entities; the votes are
simple---not ranks or scores. Let $\maxvote$ be the maximum number of votes a
single ballot can contain for that contest. We can use the above general
technique to derive an assorter for the assertion $p_A > p_B + d$. In
\autoref{sec:HesseExample} we will use this for auditing Hamiltonian free list
contests, where $A$ and $B$ will be parties. This assertion checks that the
proportion of votes $A$ has is greater than that of $B$ plus a constant, $d$.
This constant may be negative.

We start with the assertion $p_A > p_B + d$. We can rewrite this in terms of
tallies as we did in the previous examples, giving the following linear form:
\begin{align*}
       \prop{A}              &>        \prop{B}              + d                \\
\frac{\tally{A}}{\totalvote} &> \frac{\tally{B}}{\totalvote} + d                \\
      \tally{A}              &>       \tally{B}              + d \, \totalvote  \\
      \tally{A} - \tally{B}
           - d \, \totalvote &>       0.
\end{align*}
The corresponding proto-assorter is
\[ g(b) = \ballot{A} - \ballot{B} - d \cdot \ballottotal. \]
If the votes are bounded above by $\maxvote$ then this has lower bound given by
\[ g(b) \geqslant -\maxvote - d \maxvote= -\maxvote\cdot (1 + d). \]
Therefore, an assorter is given by
\[ h(b) = \frac{\ballot{A} - \ballot{B} - d \cdot \ballottotal +
                \maxvote \cdot (1 + d)}{2 \maxvote \cdot (1 + d)}. \]
When $\maxvote = 1$ this reduces to the pairwise difference assorter for
`simple' Hamiltonian contests, where each ballot can only cast a single vote
\cite{hamiltonian}. When $d = 0$ this reduces to the pairwise majority
assorter in the more general context where we can have multiple votes per
ballot.


\section{Case study: 2016 Hesse local elections}
\label{sec:HesseExample}

In the local elections in Hesse, Germany, each ballot allows the voter to cast
$\seats$ direct votes, where $\seats$ is the number of seats in the region.
Each party can have at most $\seats$ candidates on the ballot. Voters can
assign up to three votes to individual candidates; they can spread these votes
amongst candidates from different parties as they like. Voters can cross out
candidates, meaning none of their votes will flow to such candidates. Finally a
voter can select a single party. The effect of this selection is that remaining
votes not assigned to individual candidates are given to the party. At the low
level these votes are then spread amongst the candidates of the party (that
have not been crossed out) by assigning one vote to the next (uncrossed out)
candidate in the selected party, starting from the top, and wrapping around to
the top once we hit the bottom, until all the remaining votes are assigned.
Budurushi~\cite{budurushi2016} provides a detailed description of the vote
casting and vote tallying rules.\footnote{The description is based on the
(German only) official information from Hesse, see
\url{https://wahlen.hessen.de/kommunen/kommunalwahlen-2021/wahlsystem}, last
accessed 24.07.2021.}

\begin{example}
Consider a contest in a region with 12 seats, and a ballot with 4 parties. The
Greens have five candidates appearing in the order Arnold, Beatrix, Charles,
Debra, and Emma. Consider a ballot that has 3 votes assigned directly to
Beatrix, Charles crossed out, three votes assigned directly to Fox (a candidate
for another party), and the Greens party selected.

Since 6 votes are directly assigned, the Greens receive the remaining 6~votes.
We start by assigning one vote of the 6 to the top candidate, Arnold, then one
to Beatrix, none to Charles, one to Debra, one to Emma, another to Arnold, and
another to Beatrix. In total, the ballot assigns 2 votes to Arnold, 5 to
Beatrix, 1 to Debra, 1 to Emma, and 3 to Fox.
\qed
\end{example}

The social choice function involves two stages. In the first stage, the
entities we consider are the parties. This stage determines how many seats are
awarded to each party. Each party is awarded the total votes assigned on a
ballot to that party via individual candidates votes and the party selection
remainder. There is a Hamiltonian election to determine the number of seats
awarded to each party. Given $\seats$ seats in the region, we award $s_e =
\lfloor \seats \prop{e} \rfloor$ to each party $e \in \entities$. The
remaining $k = \seats - \sum_{e \in \entities} s_e$ seats are awarded to the
$k$ parties with greatest remainders $r_e = \seats \prop{e} - s_e$. Let $a_e$
be the total number of seats awarded to party $e$ (which is either $s_e$ or
$s_e + 1$).

In the second stage, seats are awarded to individual candidates. For each
party $e$ awarded $a_e$ seats, those $a_e$ candidates in the party receiving
the most votes are awarded a seat.

Performing a risk-limiting audit on a Hesse local election involves a number of
assertions. The first stage is a Hamiltonian election. The assertions required
to verify the result are described by Blom~\emph{et al.}~\cite{hamiltonian}.
For each pair of parties $m \neq n$ we need to test the assertion
\begin{equation}
  p_m > p_n + \frac{a_m - a_n - 1}{\seats}, \quad n,m \in \entities, n \neq m.
\end{equation}
While Blom~\emph{et al.}~\cite{hamiltonian} define an assorter for this
assertion, it is made under the assumption that each ballot contains a vote for
at most one entity. The assorter defined in \autoref{sec:pairdiff}---with $A =
m$, $B = n$ and $d = (a_m - a_n - 1) / S$---is more general and allows for
multiple votes per ballot.

These (All-Seats) assertions may require large samples to verify. We can
verify a simpler assertion---that each party $e$ deserved to obtain at least
$s_e$ seats---using the assertion $p_e > s_e / \seats$. We check this with an
`All-But-Remainder' audit.

The second stage of the election is a multi-winner first-past-the-post contest
within each party: party $e$'s $a_e$ seats are allocated to the $a_e$
individual candidates with highest tallies. An audit would require comparing
each winner's tally to each loser's. The margins are often very small---the
example data includes margins of only one vote---so these allocations are
likely to require a full recount, and we have not included them in our
simulations.

For experiments we consider a collection of 21 local district-based elections
held in Hesse, Germany, on March 6, 2016. An `All-But-Remainder' audit checks
that each party $e$ deserved the seats awarded to it in the first phase of
distribution ($s_e$), excluding those assigned to parties on the basis of their
`remainder'. An `All-Seats' audit checks $a_e$, i.e.\ all of the seats awarded
to party $e$, including their last seat awarded on the basis of their
remainder (if applicable).

Across the 21~district contests in our case study, the number of seats
available varied from 51 to 87, the number of parties from 6 to 11, and the
number of voters from 39,839 to 157,100. For each assertion, we estimate the
number of ballot checks required to audit it, assuming no errors are present
between each paper ballot and its electronic record. \autoref{tab:Hesse2016}
shows the number of ballot checks required to audit the most difficult
assertion in each of these contests as the contest's ASN (average sample
number) for the two levels of auditing (All-But-Remainder and All-Seats). An
ASN of $\infty$ indicates that a full manual recount would be required. We
record the ASN for risk limits, of 5\% and 10\%. The Kaplan--Kolmogorov risk
function (with $g$ = 0.1) was used to compute ASNs, given the margin for an
assertion, following the process outlined in \autoref{sec:EstimateASN}.

\autoref{tab:Hesse2016} shows that an All-Seats audit can be challenging in
terms of the sample size required, but that an All-But-Remainder audit is
usually quite practical. The estimated sample size required in an audit depends
on the margin of each assertion being checked. Where these margins are
small---for example, where two parties receive a similar remainder---the
average sample size is likely to be large. This is an inherent property of the
social choice function, not a failure of our method. For example, the All-Seats
audit for Limburg-Weilburg has an infinite ASN. The vote data shows why: the
lowest remainder to earn an extra seat is the CDU Party's, with a remainder of
24,267 votes; the highest remainder \emph{not} to earn an extra seat is the FW
Party's, with 24,205 votes. An audit would need to check that the FW did not,
in fact, gain a higher remainder than the CDU. However, a single ballot can
contain up to 71 votes, so this comparison (and hence the electoral outcome)
could be altered by a single misrecorded ballot. An electoral outcome that can
be altered by the votes on one ballot requires a full manual count in any
election system, regardless of the auditing method.

Even the All-Seats audit is quite practical when the margins represent a
relatively large fraction of ballots. This is consistent with prior work
(\cite{hamiltonian}) on US primaries, showing that an All-Seats audit is quite
practical in that context.

\subsection{Estimating an initial sample size using a risk function}
\label{sec:EstimateASN}

We use the margin of the assorter for each assertion to estimate the number of
ballot checks required to confirm that an assertion holds in an audit. As
defined in \cite{shangrla}, the margin for assertion $A$ is 2 times its
assorter mean, $\bar{h}_A$, minus 1.

Let $V$ the total number of valid ballots and $I$ be the total number of
invalid ballots cast in the contest. Note that the sum $V+I$ may differ from
the total number of votes, $\totalvote$, since there may be multiple votes
expressed on each ballot.

For an All-But-Remainder assertion indicating that party $e$ received more than
proportion $t$ of the total vote, $\totalvote$, the assorter mean is
\[
  \bar{h} = \frac{1}{V + I}
            \left(\frac{1}{2t} \tally{e} - \frac{1}{2} \totalvote +
                  \frac{1}{2} (V + I)\right),
\]
where $\tally{e}$ is the total number of votes for all candidates in party $e$.
We compute $t$ for a given assertion as follows:
\[
     q =              \frac{\totalvote}{\seats},         \quad
\delta = \left\lfloor \frac{\tally{e}}{q} \right\rfloor, \quad
     t =              \frac{q \delta}{\totalvote}.
\]
For an All-Seats comparative difference assertion between two parties, $A$ and
$B$, we need to test a pairwise difference assertion where the difference is
given by
\[
    d = \frac{(a_A - a_B - 1)}{\seats}.
\]
The assorter mean for testing this assertion is given by
\[
  \bar{h} = \frac{1}{V + I}
            \left(\frac{\tally{A} - \tally{B} - \totalvote d +
                  V \seats \cdot (1 + d)}{2 \seats \cdot (1 + d)} +
                  \frac{I}{2}\right).
\]

Once we have computed the assorter mean for an assertion, we use functionality
from the SHANGRLA software
implementation,\footnote{\texttt{TestNonnegMean.initial\_sample\_size()} from
\url{https://github.com/pbstark/SHANGRLA/blob/main/Code/assertion\_audit\_utils.py},
last accessed 24.07.2021.}
using the Kaplan--Kolmogorov risk function with $g = 0.1$, and an error rate of
0.

\begin{table}[h]
\centering
\caption{Estimates of audit sample sizes for each local district election held
in Hesse on March 6th, 2016. We record the number of assertions to be checked
in an All-But-Remainder and All-Seats audit, alongside the estimated number of
ballot checks required to complete these audits for risk limits of 5\% and
10\%, assuming no discrepancies are found between paper ballots and their
electronic records. $\seats$ is the number of seats, $|\election|$ is the total
number of ballots cast, $|\entities|$ is the total number of parties, and $V$
is the total number of valid ballots. $|\election|$ and $V$ are recorded to the
nearest thousand.}
\begin{tabular}{|l@{~}r@{~}r@{~}r@{~}r@{~}|rrr@{~}|rrr@{~}|}
\hline
    &    &    &    &    &
\multicolumn{3}{c|}{\textbf{All-But-Remainder}} &
\multicolumn{3}{c|}{\textbf{All-Seats}}  \\
\cline{6-11}
\textbf{District} & $\seats$ & $|\election|$ & $|\entities|$ & $V$ &
         & RL 5\%  & RL 10\%  &
         & RL 5\%  & RL 10\%  \\
 & & & & & $|\mathcal{A}|$ & ASN & ASN &
           $|\mathcal{A}|$ & ASN & ASN \\
\hline
Marburg-Biedenkopf   & 81 &  92k &  8 &  88k &  8 &   128 &  99 &  56 &  2,004   &  1,544   \\
Fulder               & 81 &  95k &  8 &  91k &  8 &    27 &  20 &  56 & 34,769   & 28,142   \\
Wetterau             & 81 & 122k & 11 & 115k & 11 &    26 &  20 & 110 & 12,570   &  9,790   \\
Gro{\ss} Gerau       & 71 &  85k & 11 &  80k & 11 &   291 & 224 & 110 &  7,844   &  6,101   \\
Limburg-Weilburg     & 71 &  67k &  7 &  64k &  7 &   879 & 677 &  42 & $\infty$ & $\infty$ \\
Kassel               & 81 & 100k &  7 &  95k &  7 & 1,180 & 909 &  42 &  4,580   &  3,540   \\
Darmstadt-Dieburg    & 71 & 113k &  8 & 107k &  8 &    39 &  30 &  56 & 86,480   & 76,879   \\
Bergstrasse          & 71 & 101k &  9 &  96k &  9 &    19 &  14 &  72 &  5,329   &  4,123   \\
Werra-Mei{\ss}ner    & 61 &  45k &  6 &  42k &  6 &     8 &   6 &  30 &  3,252   &  2,522   \\
Hersfeld-Rotenburg   & 61 &  52k &  7 &  50k &  7 &    29 &  23 &  42 &  5,173   &  4,026   \\
Offenbach            & 87 & 119k &  9 & 113k &  9 &    35 &  27 &  72 & 25,691   & 20,323   \\
Rheingau Taunus      & 81 &  78k &  7 &  74k &  7 &    27 &  21 &  42 &  4,382   &  3,392   \\
Lahn-Dill            & 81 &  88k &  8 &  83k &  8 &    50 &  38 &  56 &  2,752   &  2,124   \\
Waldeck-Frankenberg  & 71 &  65k &  8 &  62k &  8 &   234 & 180 &  56 &  1,508   &  1,162   \\
Main-Taunus          & 81 &  95k &  8 &  91k &  8 &    66 &  51 &  56 & 23,669   & 18,808   \\
Schwalm-Eder         & 71 &  82k &  8 &  78k &  8 &    24 &  18 &  56 & 35,724   & 29,301   \\
Odenwald             & 51 &  40k &  7 &  38k &  7 &    74 &  57 &  42 &    933   &    719   \\
Main-Kinzig          & 87 & 157k & 10 & 148k & 10 &    15 &  12 &  90 &  4,105   &  3,165   \\
Landkreis Gie{\ss}en & 81 & 103k &  8 &  98k &  8 &    41 &  24 &  56 &  8,324   &  6,464   \\
Hochtaunus           & 71 &  94k &  8 &  90k &  8 &    83 &  64 &  56 & 36,978   & 30,069   \\
Vogelsberg           & 61 &  50k &  7 &  47k &  7 &    10 &   8 &  42 &  9,668   &  7,624   \\
\hline
\end{tabular}
\label{tab:Hesse2016}
\end{table}


\section{Example: Assorters for D'Hondt and related methods}
\label{sec:dHondtExample}

Risk-limiting audits for D'Hondt and other highest averages methods were
developed by Stark and Teague~\cite{stark2014verifiable}. In this section we
show how to express those audits in the form of assertions, and develop the
appropriate assorters.

\subsection{Background on highest averages methods}

Highest averages methods are used by many parliamentary democracies in Europe,
as well as elections for the European Parliament (which uses
D'Hondt).\footnote{\url{https://www.europarl.europa.eu/RegData/etudes/BRIE/2019/637966/EPRS_BRI(2019)637966_EN.pdf},
last accessed 24.07.2021.}

Highest averages methods are similar to Hamiltonian methods in that they
allocate seats to parties in approximate proportion to the fraction of the
overall vote they won. They differ in how they allocate the last few seats when
the voting fractions do not match an integer number of seats.

A highest averages method is parameterized by a set of divisors $d(1), d(2),
\dots$ $d(\seats)$ where $\seats$ is the number of seats. The seats are
allocated by forming a table in which each party's votes are divided by each of
the divisors, then choosing the $\seats$ largest numbers in the whole
table---the number of selected entries in a party's row is the number of seats
that party wins. The divisors for D'Hondt are $d(i) = i$, $i = 1, 2, \dots
\seats$. Sainte-Lagu\"{e} has divisors $d(i) = 2i - 1$, for $i = 1,  2, \dots
\seats$.

Let $f_{e,s} = \tally{e} / d(s)$ for entity $e$ and seat $s$. The
\emph{Winning Set} $\mathcal{W}$ is
\[
  \mathcal{W} = \{ (e,s) : f_{e,s} \text{ is one of the $\seats$ largest} \}.
\]
This can be visualised in a table by writing out, for each entity $e$, the
sequence of numbers $T_e/d(1), T_e/d(2), T_e/d(3), \ldots$, and then selecting
the $\seats$ largest numbers in the table. Each party receives a number of
seats equal to the number of selected values in its row.

%
%

Like Hamiltonian methods, highest averages methods can be used in a simple form
in which voters choose only their favourite party, or in a variety of more
complex forms in which voters can express approval or disapproval of individual
candidates. We deal with the simple case first.

\subsection{Simple D'Hondt: Party-only voting}

In the simplest form of highest averages methods, seats are allocated to each
entity (party) based on individual entity tallies. Let $W_e$ be the number of
seats won and $L_e$ the number of the first seat lost by entity $e$. That is:
\begin{align*}
W_e &= \max \{ s : (e,s) \in    \mathcal{W} \}; \perp \text{if $e$ has no winners.} \\
L_e &= \min \{ s : (e,s) \notin \mathcal{W} \}; \perp \text{if $e$ won all the seats.}
\end{align*}
If $e$ won some, but not all, seats, then $L_e = W_e + 1$.

The inequalities that define the winners are, for all parties $A$ with at least
one winner, for all parties $B$ (different from $A$) with at least one loser,
as follows:
\begin{equation}
\label{eq:DHondtTest}
  f_{A,W_A} > f_{B,L_B}.
\end{equation}
Converting this into the notation of \autoref{sec:GeneratingAssorters},
expressing \autoref{eq:DHondtTest} as a linear assertion gives us, $\forall A
\textit{ s.t. } W_A \neq \perp, \forall B \neq A \textit{ s.t. } L_B \neq
\perp,$
\[
  \tally{A} / d(W_A) - \tally{B} / d(L_B) > 0.
\]
From this, we define the proto-assorter for any ballot $b$ as
\[
  g_{A,B}(b) :=
  \begin{cases}
    \phantom{-} 1 / d(W_A) & \text{if $b$ is a vote for party $A$,} \\
             -  1 / d(L_B) & \text{if $b$ is a vote for party $B$,} \\
    \phantom{-} 0          & \text{otherwise,}
  \end{cases}
\]
\[
  \text{or equivalently} \quad g_{A,B}(b) := \ballot{A} / d(W_A) -
                                             \ballot{B} / d(L_B)
\]
where $\ballot{A}$ (resp.\ $\ballot{B}$) is 1 if there is a vote for party $A$
(resp.\ $B$), 0 otherwise.

\noindent
The lower bound is clearly $a = -1 / d(L_B)$. Substituting into
\autoref{eq:assorter} gives
\begin{align*}
h_{A,B}(b)
  &= \begin{cases}
      1/2 \left[d(L_B) / d(W_A) + 1\right] & \text{if $b$ is a vote for party $A$,} \\
        0                                  & \text{if $b$ is a vote for party $B$,} \\
      1/2                                  & \text{otherwise.}
     \end{cases}
\end{align*}
Note that order matters: in general, both $h_{A,B}$ and $h_{B,A}$ are
necessary---the first checks that party $A$'s lowest winner beat party $B$'s
highest loser; the second checks that party $B$'s lowest winner beat party
$A$'s highest loser.

\subsection{More complex methods: Multi-candidate voting}

Like some Hamiltonian elections, many highest averages elections also allow
voters to select individual candidates. A party's tally is the total of its
candidates' votes. Then, within each party, the won seats are allocated to the
candidates with the highest individual tallies. The main entities are still
parties, allocated seats according to \autoref{eq:DHondtTest}, but the assorter
must be generalised to allow one ballot to contain multiple votes for various
candidates.

The proto-assorter for entities (parties) $A \neq B \textit{ s.t. } W_A \neq
\perp, \text{ and } L_B \neq \perp,$ is very similar to the single-party case,
but votes for each party ($\ballot{A}$ and $\ballot{B}$) count the total, over
all that entity's candidates, and may be larger than one.
\[
  g_{A,B}(b) := \ballot{A} / d(W_A) - \ballot{B} / d(L_B).
\]
The lower bound is $-\maxindvote / d(L_B)$, again substituting in to
\autoref{eq:assorter} gives
\begin{align*}
h_{A,B}(b)
  &= \frac{\ballot{A} d(L_B) / d(W_A) - \ballot{B} + \maxindvote}{2\maxindvote}.
\end{align*}
Note this reduces to the single-vote assorter when $\maxindvote = 1$
($\ballot{A}, \ballot{B} \in \{0,1\}$).


\section{Conclusion and future work}

SHANGRLA reduces RLAs for many social choice functions to a canonical form
involving `assorters.' This paper shows how to translate general linear
assertions into canonical assorter form for SHANGRLA, illustrated by developing
the first RLA method for Hamiltonian free list elections and the first
assertion-based approach for D'Hondt style elections.

We show that party-list proportional representation systems can be audited
using simple assertions that are both necessary and sufficient for the reported
outcome to be correct. In some settings, including in Hesse, elections are
inherently expensive to audit because margins are frequently small, both
between parties vying for the seats allocated by remainder, and between
candidates in the same party.

There are social choice functions for which no set of linear assertions
guarantees the reported winner really won, for instance, social choice
functions in which the order of in which the votes are tabulated matters or
that involve a random element. Some variants of Single Transferable Vote (STV)
have one or the other of those properties.

Other variants of STV might be amenable to RLAs and to SHANGRLA in particular:
the question is open. We conjecture that STV is inherently hard to audit.
Although a sufficient set of conditions is easy to generate---simply check
every step of the elimination and seat-allocation sequence---this is highly
likely to have very small margins and hence to require impractical sample
sizes. We conjecture that it is hard to find a set of conditions that imply an
STV outcome is correct and that requires reasonable sample sizes to audit. Of
course, this was also conjectured for IRV and turns out to be false.


\bibliographystyle{splncs04}
\bibliography{main}


\end{document}